\documentclass[preprintnumbers,amsmath,amssymb,12pt,floatfix,epsfig]{revtex4}
\usepackage{dcolumn}
\usepackage{bm}
\usepackage[dvips]{color}
\usepackage{graphicx}
\leftmargin -0.5in \baselineskip=24ptplus.5ptminus.2pt \vspace*{0.5
in} \large
\begin{document}
\title{Extended Optical Model Analyses of
Elastic Scattering and Fusion Cross Section Data for
the $^{12}$C+$^{208}$Pb System at Near-Coulomb-Barrier Energies
by using a Folding Potential}
\vspace{0.5cm}

\author{W. Y. So and T. Udagawa}
\address{Department of Physics, University of Texas,
Austin, Texas 78712}

\author{S. W. Hong and B. T. Kim}
\affiliation{\it Department of Physics and Institute of Basic Science, \\
Sungkyunkwan University, Suwon 440-746, Korea}

\begin{abstract}
\begin{center}
{\vspace{1.0cm} \bf Abstract}
\end{center}
Simultaneous $\chi^{2}$ analyses are performed for elastic
scattering and fusion cross section data for the $^{12}$C+$^{208}$Pb
system at near-Coulomb-barrier energies by using the extended
optical model approach in which the polarization potential is
decomposed into direct reaction (DR) and fusion parts.
Use is made of the double folding potential as a bare potential.  It
is found that the experimental elastic scattering and fusion data
are well reproduced without introducing any normalization factor for
the double folding potential and also that both DR and fusion parts
of the polarization potential determined from the $\chi^{2}$
analyses satisfy separately the dispersion relation.
Furthermore, it is shown that the imaginary parts of both DR and fusion
potentials at the strong absorption radius change very rapidly,
which results in a typical threshold anomaly
in the total imaginary potential
as observed with tightly bound projectiles
such as $\alpha$-particle and $^{16}$O.
\end{abstract}
\maketitle \vspace{1.5cm} PACS numbers : 24.10.-i,~25.70.Jj

\pagebreak

\section{Introduction}

Recently we carried out analyses~\cite{so1,so2,so3} based on
the extended optical model~\cite{uda1,hong,uda2}, in which the
optical potential consists of the energy independent Hartree-Fock
potential and the energy dependent complex polarization potential
that has two components, i.e., the direct reaction (DR) and fusion
parts, which we call the DR and fusion potentials, respectively. In
the original work based on the extended optical
model~\cite{uda1,hong,uda2}, use was made of a usual
Woods-Saxon potential for the Hartree-Fock part of the potential, but
in Refs.~\cite{so1,so2,so3}, we started using the double folding
potential~\cite{sat1}.

The main interest in the studies of Refs.~\cite{so1,so2,so3} was
the normalization constant $N$ introduced earlier to reproduce
the elastic scattering data for loosely bound
projectiles such as $^{6}$Li and $^{9}$Be; in the analysis of
data for such loosely bound projectiles using the usual
optical model with a folding potential~\cite{sat1}
one was forced to reduce the
strength of the folding potential by a factor $N=0.5 \sim 0.6$ in order to
reproduce the data. This reduction factor was later ascribed
to the strong breakup character of the projectiles.
Studies were made on the effects of the breakup on the
elastic scattering, based on the coupled discretized continuum
channel (CDCC) method~\cite{sak1,kee2}. These studies were very
successful in reproducing the elastic scattering data without
introducing any arbitrary normalization factors and further in
understanding the physical origin of the factor $N=0.5 \sim 0.6$
needed in one channel optical model
calculations. The authors of Refs.~\cite{sak1,kee2}
projected their coupled channel equations to a single elastic
channel equation and deduced the polarization potential arising from
the coupling with the breakup channels. The resultant real part of
the polarization potential was then found to be repulsive at the
surface region around the strong absorption radius, $R_{sa}$. This
means that the reduction of the folding potential by a factor of
$N=0.5 \sim 0.6$ needed in the one-channel
optical model calculation is to effectively take into
account the effects of the coupling with the breakup channels.

We explored this problem for the $^{6}$Li~\cite{so1},
$^{7}$Li~\cite{so2} and $^{9}$Be~\cite{so3} induced scattering and
fusion in the framework of the extended optical model with the
double folding potential. Simultaneous $\chi^{2}$ analyses of the
elastic scattering and fusion cross section data were performed 
to determine the two types of the polarization potentials as
functions of the incident energy $E_{lab}$. Our expectation was that
the resulting real part of the DR potential would become repulsive
consistently with the results of the CCDC calculations. We have
indeed obtained repulsive real DR polarization
potentials~\cite{so1,so2,so3}. In addition, it was shown that both
DR and fusion potentials satisfied the dispersion
relation~\cite{mah1,nag1} separately.

In the present study, we extend the work of Refs.~\cite{so1,so2,so3} to the
$^{12}$C+$^{208}$Pb system. Since $^{12}$C is a tightly bound projectile,
such an anomalous normalization constant $N = 0.5 \sim 0.6$
observed in $^{6}$Li or $^{9}$Be induced scattering is not expected
around the Coulomb barrier energies. In fact, the normalization
factor $N$ for reproducing the $^{12}$C projectile
data was found to be close
to unity, $N \approx 1$ (see Ref.~\cite{sat1}).

In Sec. II, we first discuss characteristic features
of elastic scattering cross section data of
$^{12}$C+$^{208}$Pb~\cite{san1} in comparison with those of $^{6}$Li
induced scattering. From this comparison
it will be shown that the DR cross section is expected to be
significantly smaller in
$^{12}$C+$^{208}$Pb than in $^{6}$Li or $^{9}$Be induced reactions.
In Sec. III, we then generate the so-called semi-experimental DR
cross section, $\sigma_{D}^{\textrm{semi-exp}}$, by using the elastic
scattering data together with the fusion cross section
data~\cite{muk1} by following the method described in, e.g.,
Ref.~\cite{so4}. The data of $\sigma_{D}^{\textrm{semi-exp}}$ is
needed in making the separate determination of the DR and fusion
potentials in the extended optical model.
Simultaneous $\chi^{2}$ analyses of the data of the elastic
scattering, fusion and semi-experimental DR cross sections are then
carried out in Sec. IV, where the results are also presented.
Section V concludes the paper.

\section{Review of Experimental Cross Sections}

We begin with the discussion of some characteristic features of the
elastic scattering cross section $d\sigma_{el}/d\sigma_{\Omega}$
data of $^{12}$C+$^{208}$Pb in comparison with those of $^{6}$Li, $^{7}$Li and
$^{9}$Be induced scattering~\cite{so1,so2,so3}.
Such features can best be seen in the ratio, $P_{E}$, defined by
\begin{equation}\label{p_e}
P_{E} \equiv \frac{d\sigma_{el}}{d\sigma_{\Omega}}/
             \frac{d\sigma_{C}}{d\sigma_{\Omega}}=d\sigma_{el}/d\sigma_{C}
\end{equation}
as a function of the distance of the closest approach $D$ (or the reduced
distance $d$), where $d\sigma_{C}/d\sigma_{\Omega}$ is the Coulomb
scattering cross section, while  $D$ ($d$) is related to the scattering angle
$\theta$ by
\begin{equation}
D=d(A_{1}^{1/3}+A_{2}^{1/3})=\frac{1}{2}D_{0}
                           (1+\frac{1}{\mbox{sin}(\theta/2)})
\end{equation}
with
\begin{equation}
D_{0}=\frac{Z_{1}Z_{2}e^{2}}{E}
\end{equation}
being the distance of the closest approach in a head-on
collision. Here $(A_{1},Z_{1})$ and $(A_{2},Z_{2})$ are the mass and
charge of the projectile and target ions, respectively, and $E
\equiv E_{c.m.}$ is the incident energy in the center-of-mass
system. $P_{E}$ as defined by Eq.~(\ref{p_e}) will be referred to as
the elastic probability.

\begin{figure}
\begin{center}
\vspace*{1.0cm}
\includegraphics[width=0.95\linewidth] {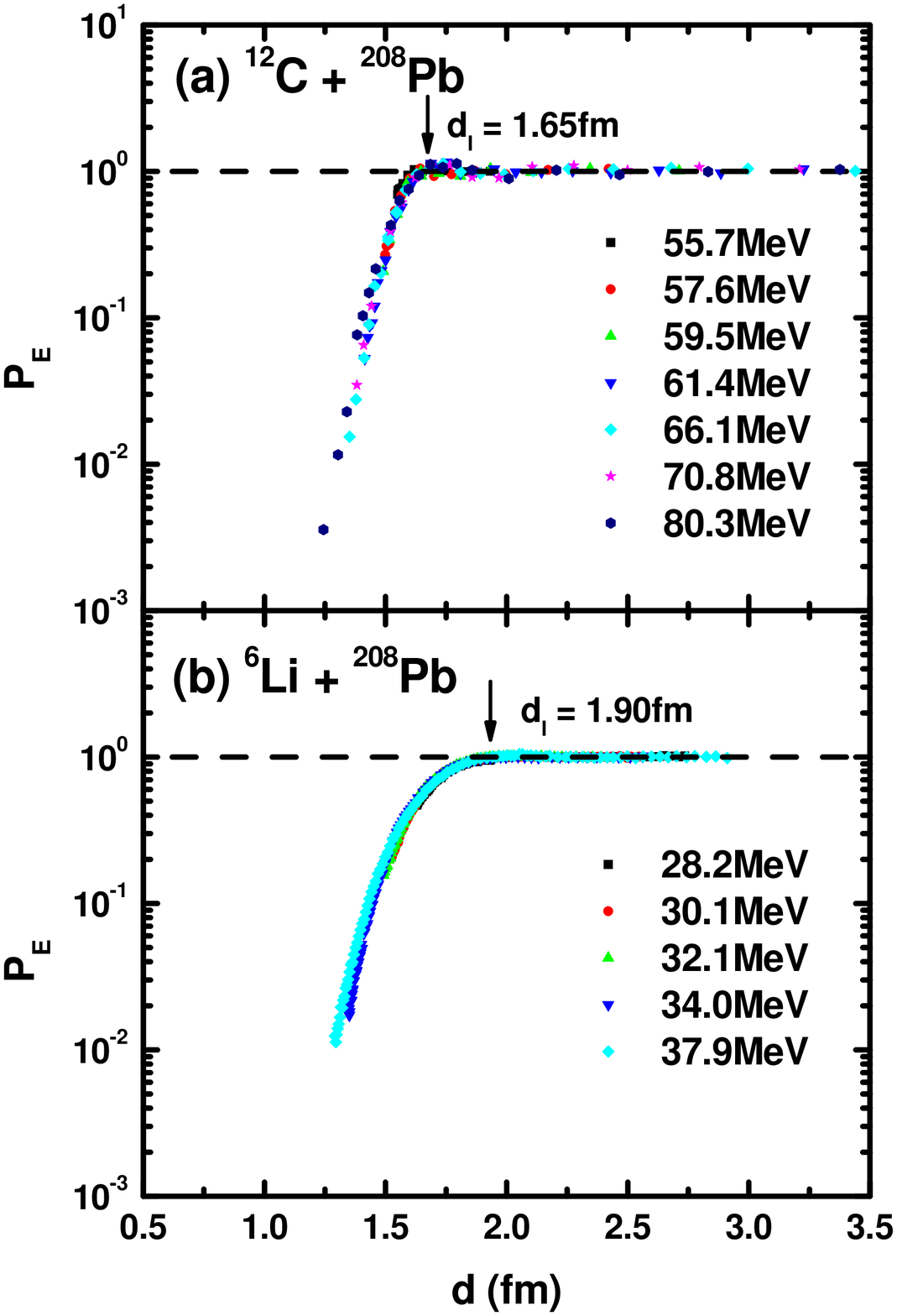}
\end{center}
\caption{\label{pe-value}~(Color online) $P_{E}$ values for the (a)
$^{12}$C+$^{208}$Pb system and (b) $^{6}$Li+$^{208}$Pb system.}
\end{figure}

In Figs.~\ref{pe-value} (a) and~\ref{pe-value} (b), we present
the experimental values of $P_{E}$ for incident energies available
around the Coulomb barrier energy as a function of the reduced
distance $d$ for $^{12}$C+$^{208}$Pb~\cite{muk1} and
$^{6}$Li+$^{208}$Pb~\cite{kee1} systems, respectively. The latter case is
presented as an example of $P_{E}$ for a loosely bound
projectile.  As seen, the values of $P_{E}$ at different energies
line up to form a very narrow band for both cases. This is a
characteristic feature observed in many of the heavy-ion collisions,
reflecting the semiclassical nature of the collisions. $P_{E}$
remains close to unity until the two ions approach each other at around
$d \sim d_{I}$, where $P_{E}$ begins to fall off. The distance
$d_{I}$ is usually called the interaction distance, at which the
nuclear interactions between the colliding ions are switched on, so
to speak. The values of $d_{I}$ are about 1.65~fm for
$^{12}$C+$^{208}$Pb and 1.9~fm for $^{6}$Li+$^{208}$Pb.

The fall off of the $P_{E}$ value in the region immediately next to
$d_{I}$ is due to DR. The fact that the $d_{I}$-value of 1.9~fm for
$^{6}$Li+$^{208}$Pb is much larger than $d_{I}=1.65$~fm for
$^{12}$C+$^{208}$Pb implies that DR starts to take place in
$^{6}$Li+$^{208}$Pb at larger distances than it does in
$^{12}$C+$^{208}$Pb. Further it is seen that the amount of decrease
of the $P_{E}$ value from unity in $^{6}$Li+$^{208}$Pb is
significantly larger than in $^{12}$C+$^{208}$Pb in the region of
$d=1.5 \sim 1.9$~fm, where DR takes place. These features clearly
indicate that DR (which may be dominated by breakup) takes place
significantly more strongly in $^{6}$Li+$^{208}$Pb than in
$^{12}$C+$^{208}$Pb.  This is indeed the case as seen from the
semi-experimental DR cross section extracted in the next
section.

Finally, we note that in the region of $d < 1.5$~fm where fusion
dominates, the values of $P_{E}$ for $^{12}$C+$^{208}$Pb and
$^{6}$Li+$^{208}$Pb become almost identical, implying that there
may not be much difference in the absorption rates of
both systems when these colliding nuclei approach each other
as close as $d < 1.5$~fm.
In fact, the $P_{E}$ value becomes 0.1 for both cases
at approximately $d=1.43 \sim 1.44$~fm.

\section{Extracting semi-experimental DR cross section}

For our purpose of determining the fusion and DR
potentials separately, it is desirable to have the data of DR cross
sections in addition to fusion and elastic scattering cross
sections. For the $^{12}$C+$^{208}$Pb system, however, no reliable
data of the DR cross section are available, although some efforts
have been devoted to measure the inelastic and transfer reaction
cross sections~\cite{san1}. Here, we thus generate the so-called
semi-experimental DR cross section $\sigma_{D}^{\textrm{semi-exp}}$,
following the method proposed in Ref.~\cite{so4}.

Our method to generate $\sigma_{D}^{\textrm{semi-exp}}$ resorts to
the well known empirical fact that the total reaction cross section
$\sigma_{R}$
calculated from the optical model fit to the available elastic
scattering cross section data, $d\sigma^{\textrm{exp}}_{E}/d\Omega$,
usually agrees well with the experimental $\sigma_{R}$, in spite of
the well known ambiguities in the optical potential. Let us call
$\sigma_{R}$ thus generated the semi-experimental reaction cross
section $\sigma_{R}^{\textrm{semi-exp}}$. Then,
$\sigma_{D}^{\textrm{semi-exp}}$ can be generated by
\begin{equation} \label{sig_D}
\sigma^{\textrm{semi-exp}}_{D} = \sigma^{\textrm{semi-exp}}_{R} -
\sigma^{\textrm{exp}}_{F}.
\end{equation}
This approach seems to work even for loosely bound projectiles, as
demonstrated by Kolata {\it et al.}~\cite{kol1} for the
$^{6}$He+$^{209}$Bi system.

Following Ref.~\cite{so4}, we first carry out rather simple optical
model $\chi^{2}$ analyses of elastic scattering data solely for the
purpose of deducing $\sigma_{R}^{\textrm{semi-exp}}$. For
these preliminary analyses, we assume the optical potential to be
the sum of $V_{0}(r)$+$i W_{I}(r)$ and $U_{1}(r,E)$, where $V_{0}(r)$ is
the real, energy independent bare folding potential discussed
in Sec. IV. B, $i W_{I}(r)$ is an energy independent short
range imaginary potential discussed in Sec. IV. A, and
$U_{1}(r,E)$ is a Woods-Saxon type complex potential with common
geometrical parameters for both real and imaginary parts. The
elastic scattering data are then fitted with a fixed radius
parameter $r_{1}$ for $U_{1}(r,E)$, treating, however, three
other parameters, the real and the imaginary strength $V_{1}$ and
$W_{1}$ and the diffuseness parameter $a_{1}$, as adjustable. The
$\chi^{2}$ fitting is done for three choices of the radius
parameter; $r_{1}$=1.3, 1.4, and 1.5 fm. These different choices of
the $r_{1}$-value are made to examine the dependence of the
resulting $\sigma_{R}^{\textrm{semi-exp}}$ on the value of $r_1$.

As observed in Ref.~\cite{so4}, the values of
$\sigma_{R}^{\textrm{semi-exp}}$ thus extracted for three different
$r_{1}$ values agree with the average value of
$\sigma_R^{\textrm{semi-exp}}$ within 3\%, implying that
$\sigma_{R}^{\textrm{semi-exp}}$ is determined without much
ambiguity. We then identified the average values as the final values
of $\sigma_{R}^{\textrm{semi-exp}}$. Using thus determined
$\sigma_{R}^{\textrm{semi-exp}}$, we generated
$\sigma_{D}^{\textrm{semi-exp}}$ by employing Eq.~(\ref{sig_D}). The
resultant values of $\sigma_{R}^{\textrm{semi-exp}}$ and
$\sigma_{D}^{\textrm{semi-exp}}$ are presented in
Table~\ref{semiexp}, together with $\sigma_{F}^{\textrm{exp}}$. In
Table~\ref{semiexp}, given are also $\sigma_{R}^{\textrm{semi-exp}}$
determined in Ref.~\cite{san1} from the optical model calculations.
The two sets of $\sigma_{R}^{\textrm{semi-exp}}$
determined independently agree within 6\% except for
the lowest energy case of $E_{cm}=$55.7~MeV where the discrepancy
amounts to 25\%. However, at this energy the value of the cross section
is very small, and thus $\sigma_{R}^{\textrm{semi-exp}}$
determined from the elastic scattering data has a relatively large
uncertainty.

\begin{table}
\caption{Semi-experimental total reaction and DR cross sections for
the $^{12}$C+$^{208}$Pb system.} \vspace{2ex} \label{semiexp}
\begin{ruledtabular}
\begin{tabular}{cccccc}
$E_{lab} $ &$E_{c.m.}$ & $\sigma^{\textrm{exp}}_{F}$ &
$\sigma^{\textrm{semi-exp}}_{D}$ & $\sigma^{\textrm{semi-exp}}_{R}$
& $\sigma^{\textrm{semi-exp}}_{R}$~\cite{san1} \\
(MeV) & (MeV) & (mb) & (mb) & (mb) & (mb) \\ \hline
58.9 & 55.7 &  14 &   1 &  15 &  20 \\
60.9 & 57.6 &  85 &  57 & 142 & 136 \\
62.9 & 59.5 & 189 & 111 & 300 & 286 \\
64.9 & 61.4 & 291 & 129 & 420 & 429 \\
69.9 & 66.1 & 520 & 179 & 699 & 715 \\
74.9 & 70.8 & 718 & 241 & 959 & 969 \\
84.9 & 80.3 &1045 & 327 &1371 &1373 \\
\end{tabular}
\end{ruledtabular}
\end{table}

\begin{figure}
\begin{center}
\vspace*{5.0cm}
\includegraphics[width=0.95\linewidth] {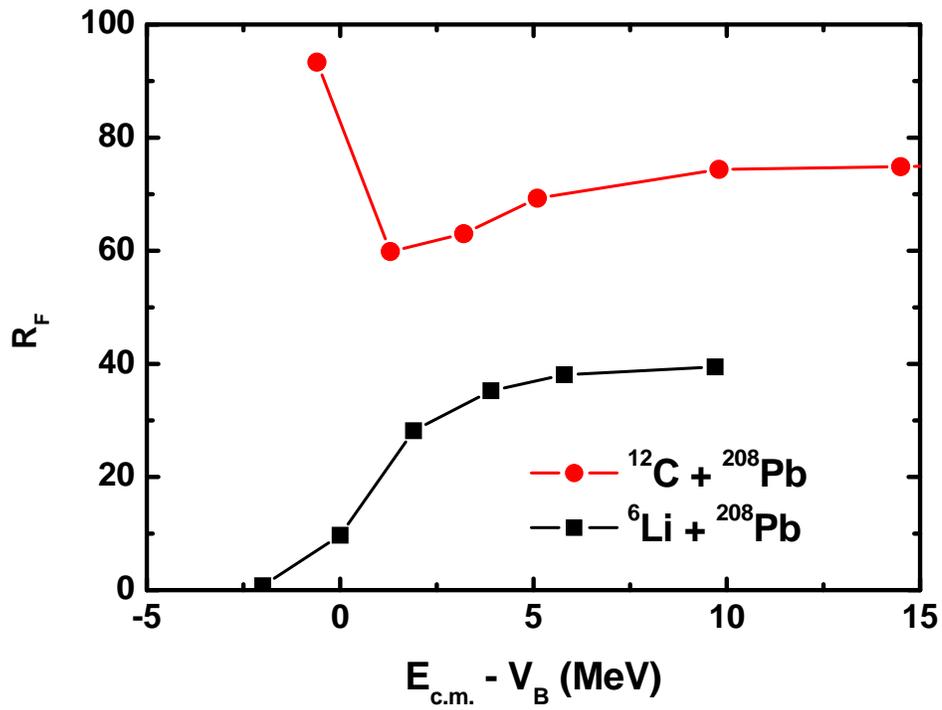}
\end{center}
\caption{\label{rf-value}~(Color online) $R_{F}\equiv
(\sigma_{F}^{exp}/\sigma_{R}^{semi-exp}) \times 100$ (\%) values for
the $^{12}$C+$^{208}$Pb and $^{6}$Li+$^{208}$Pb systems.}
\end{figure}

As seen in Table~\ref{semiexp}, $\sigma_{F}^{\textrm{exp}}$ is much
larger than $\sigma_{D}^{\textrm{semi-exp}}$, and
$\sigma_{R}^{\textrm{semi-exp}}$ is dominated by
$\sigma_{F}^{\textrm{exp}}$. This is quite in contrast to the case
for the $^{6}$Li+$^{208}$Pb system, where
$\sigma_{R}^{\textrm{semi-exp}}$ is dominated by
$\sigma_{D}^{\textrm{semi-exp}}$ (see Table I of Ref.~\cite{so1}. To
demonstrate differences, we present in Fig.~\ref{rf-value} the
ratio, $R_{F}$, defined by

\begin{equation}
R_{F} \equiv
(\sigma_{F}^{\textrm{exp}}/\sigma_{R}^{\textrm{semi-exp}}) \times
100
\end{equation}
for both $^{12}$C+$^{208}$Pb and $^{6}$Li+$^{208}$+Pb systems as a
function of $E_{cm}-V_{B}$, $V_{B}$ being the Coulomb barrier
height.  It is seen that the $R_{F}$ values for the
$^{12}$C+$^{208}$Pb system are larger than 50\% at all the energies
considered and become close to 100\% at $E_{cm} < V_{B}$, while for
the $^{6}$Li+$^{208}$Pb system the $R_F$ values are less than 50\% everywhere
and become close to zero at $E_{cm} < V_{B}$. For the
$^{12}$C+$^{208}$Pb system, the reaction is dominated by
fusion, in particular near and below the Coulomb barrier energy.

\section{Simultaneous $\chi^{2}$ Analyses}

Simultaneous $\chi^{2} $analyses were then performed on the data sets of
($d\sigma^{\textrm{exp}}_{E}/d\Omega$,~$\sigma_{D}^{\textrm{semi-exp}}$,
~$\sigma^{\textrm{exp}}_{F}$) by taking the data for
$d\sigma^{\textrm{exp}}_{E}/d\Omega$, and
$\sigma^{\textrm{exp}}_{F}$ from the literature~\cite{san1,muk1}. In
calculating the $\chi^{2}$ value, we simply assumed 1\% errors for
all the experimental data. The 1\% error is about the average of
errors in the measured elastic scattering cross sections, but much
smaller than the errors in the DR ($\sim$5\%) and fusion
($\sim$10\%) cross sections. The choice of the 1\% error for DR and
fusion cross sections is thus equivalent to increasing the weight
for the DR and fusion cross sections
by factors of 25 and 100, respectively. Such a
choice of errors may be reasonable, since we have only one datum
point for each of these cross sections, while there are more than
10 data points for the elastic scattering cross sections.

\subsection{Necessary Formulas}

The optical potential $U(r,E)$ we use in the present work has the
following form;
\begin{equation} \label{u}
U(r;E) = V_{C}(r)-[V_{0}(r)+U_{F}(r;E)+U_{D}(r;E)],
\end{equation}
where $V_{C}(r)$ is the usual Coulomb potential with $r_{C}$=1.25 fm
and $V_{0}(r)$ is the bare nuclear potential, for which use is made
of the double folding potential described in
the next subsection. $U_{F}(r;E)$ and $U_{D}(r;E)$ are,
respectively, fusion and DR parts of the so-called polarization
potential~\cite{love} that originates from couplings to the
respective reaction channels. Both $U_{F}(r;E)$ and $U_{D}(r;E)$ are
complex and their forms are assumed to be of volume-type and
surface-derivative-type~\cite{kim1,hong}, respectively.
$U_{F}(r;E)$, and $U_{D}(r;E)$ are explicitly given by
\begin{equation} \label{u_f}
U_{F}(r;E) = (V_{F}(E)+iW_{F}(E))f(X_{F})+iW_{I}(r),
\end{equation}
and
\begin{equation} \label{u_d}
U_{D}(r;E) = (V_{D}(E)+iW_{D}(E))4a_{D}\frac{df(X_{D})}{dR_{D}},
\vspace{2ex}
\end{equation}
where $f(X_{i})=[1+\mbox{exp}(X_{i})]^{-1}$ with
$X_{i}=(r-R_{i})/a_{i}$ $({\it i}=F\; \mbox{and} \; D)$ is the usual
Woods-Saxon function with the fixed geometrical parameters of
$r_{F}=1.40$~fm, $a_{F}=0.35$~fm, $r_{D}=1.50$~fm, and
$a_{D}=0.55$~fm, while $V_{F}(E)$, $V_{D}(E)$, $W_{F}(E)$, and
$W_{D}(E)$ are the energy-dependent strength parameters. Since we
assume the geometrical parameters of the real and imaginary
potentials to be the same, the strength parameters $V_{i}(E)$ and
$W_{i}(E)$ ($i=F$ or $D$) are related through a dispersion
relation~\cite{mah1},
\begin {equation} \label{disper}
V_{i}(E)=V_{i}(E_{s}) + \frac {E-E_{s}}{\pi } \mbox{P}
\int_{0}^{\infty} dE' \frac {W_{i}(E')}{(E'-E_{s})(E'-E)},
\vspace{2ex}
\end {equation}
where P stands for the principal value and $V_{i}(E_{s})$ is the
value of $V_{i}(E)$ at a reference energy $E=E_{s}$. Later, we will
use Eq.~(\ref{disper}) to generate the final real strength
parameters $V_{F}(E)$ and $V_{D}(E)$ using $W_{F}(E)$ and $W_{D}(E)$
fixed from $\chi^{2}$ analyses.

The last imaginary potential $W_{I}(r)$ in $U_{F}(r;E)$ given by
Eq.~(\ref{u_f}) is a short-range potential of the Woods-Saxon type
given by
\begin{equation}
W_{I}(r) = W_{I}f(X_{I}),
\end{equation}
with $W_{I}=40$~MeV, $r_{I}=1.0$~fm, and $a_{I}=0.30$~fm.
This imaginary potential is introduced to eliminate
unphysical oscillations appearing in the radial wave functions of
low partial waves when this $W_{I}(r)$ is not included. Because of the
deep nature of the folding potential $V_{0}$ used in this study and
also because $W_{F}(E)f(X_{F})$, another imaginary part in
$U_{F}(r;E)$, turns out to be not strong enough, reflections of lower
partial waves appear, which causes the oscillations mentioned above,
but physically such oscillations should not occur. $W_{I}(r)$ is introduced in
order to eliminate this unphysical effect. We might introduce
a real part $V_{I} (r)$ associated with $W_I (r)$,
but we ignore this part, simply
because such a real potential does not affect at all real
physical observables, which means that it is impossible to extract
the information of $V_I (r)$ from analyzing the experimental data.

In the extended optical model,
fusion and DR cross sections, $\sigma^{\textrm{th}}_{F}$
and $\sigma^{\textrm{th}}_{D}$,
respectively,  are calculated by using the following
expression~\cite{uda1,hong,uda2,huss}
\begin {equation}
\sigma^{\textrm{th}}_{i} = \frac {2}{\hbar v} <\chi^{(+)}|
\mbox{Im}~[U_{i}(r;E)]|~\chi^{(+)}> \hspace{.5in}
(i=F\;\mbox{or}\;D),
\end{equation}
where $\chi^{(+)}$ is the usual distorted wave function that
satisfies the Schr\"{o}dinger equation with the full optical model
potential $U(r;E)$ in Eq.~(\ref{u}). $\sigma^{\textrm{th}}_{F}$ and
$\sigma^{\textrm{th}}_{D}$ are thus calculated within the same
framework as $d\sigma_{E}/d\Omega$ is calculated. Such a unified
description enables us to evaluate all the different types of cross
sections on the same footing.

\subsection{The Folding Potential}

The double folding potential $V_{0}(r)$ we use in the present study
as the bare potential may be written as~\cite{sat1}
\begin{equation}
V_{0}(r)=\int d{\bf r}_{1} \int d{\bf r}_{2} \rho_{1}(r_{1})
\rho_{2}(r_{2}) v_{NN}(r_{12}=|\bf{r}-\bf{r}_{1}+\bf{r}_{2}|),
\end{equation}
where $\rho_{1}(r_{1})$ and $\rho_{2}(r_{2})$ are the nuclear matter
distributions for the target and projectile nuclei, respectively,
while $v_{NN}$ is the M3Y interaction that describes the
effective nucleon-nucleon interaction and the knockon exchange
effect given as
\begin{equation}
v_{NN}(r)=7999\frac{e^{-4r}}{4r}-2134\frac{e^{-2.5r}}{2.5r}-262
\delta (r).
\end{equation}
For $\rho_{1}(r)$ we use the following Woods-Saxon form taken from
Ref.~\cite{jag1}:
\begin{equation}
\rho_{1}(r)=\rho_{0}/\left[1+\mbox{exp}\left(\frac{r-c}{z}\right)\right],
\end{equation}
with $c=6.624$~fm and $z=0.549$~fm, while for $\rho_{2}(r)$ the
following form is taken from Ref.~\cite{jag1}:
\begin{equation}
\rho_{2}(r)=\rho_{0} (1+wr^{2}/c^{2})/\left[1+
\mbox{exp}\left(\frac{r-c}{z}\right)\right],
\end{equation}
with $c=2.355$~fm, $z=0.522$~fm, and $w=-0.149$~fm. We then use the code
DFPOT of Cook~\cite{coo1} for evaluating $V_{0} (r)$.

\subsection{Threshold Energies of Subbarrier Fusion and DR}

As in Ref.~\cite{so1}, we utilize as an important ingredient the
so-called threshold energies $E_{0,F}$ and $E_{0,D}$ of subbarrier
fusion and DR, respectively, which are defined as zero intercepts of
the linear representation of the quantities $S_{i}(E)$, defined by
\begin{equation}\label{s_i}
S_{i} \equiv \sqrt{E \sigma_{i}} \approx \alpha_{i} (E-E_{0,i})
\;\;\; (i=F \; \mbox{or} \; D),
\end{equation}
where $\alpha_{i}$ is a constant. $S_{i}$ with $i=F$, i.e., $S_{F}$
is the quantity introduced originally by Stelson {\it et
al.}~\cite{stel}, who showed that in the subbarrier region $S_{F}$
from the measured $\sigma_{F}$ could be represented very well by a
linear function of $E$ (linear systematics) as in Eq.~(\ref{s_i}).
In Ref.~\cite{kim1}, we extended the linear systematics to DR cross
sections. In fact the DR data are also well represented by a linear
function.

\begin{figure}
\begin{center}
\vspace*{5.0cm}
\includegraphics[width=0.95\linewidth] {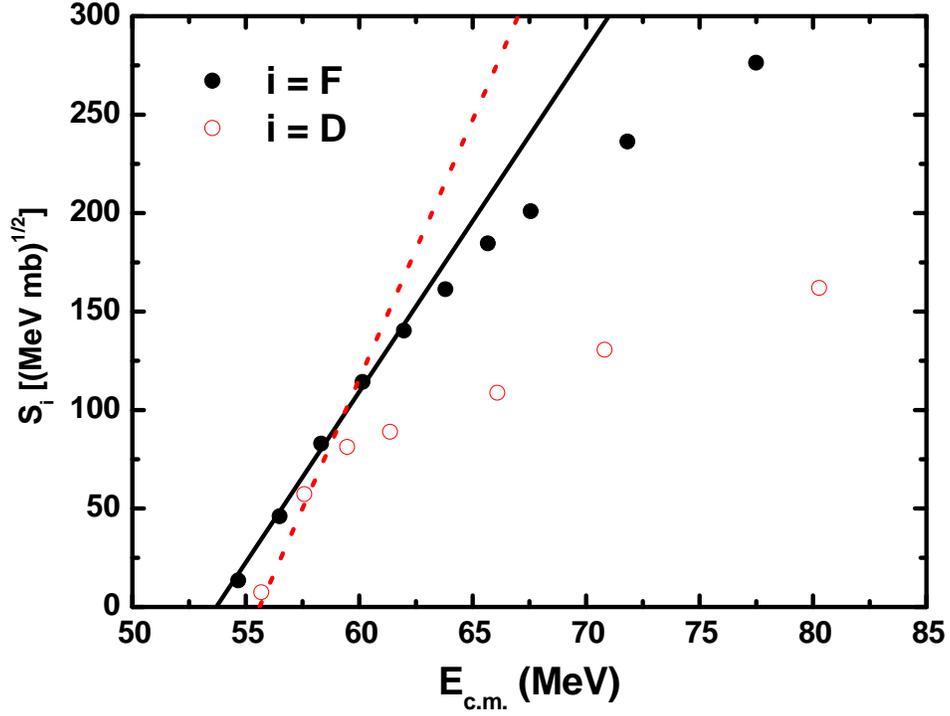}
\end{center}
\caption{\label{s-factor}~(Color online) The Stelson plot of
$S_{i}=\sqrt{E~ \sigma_{i}}$ for DR ($i=D$, open circles) and fusion
($i=F$, filled circles) cross sections. Use is made of the
semi-experimental DR cross section for $S_{D}$, while the
experimental fusion cross section is employed for $S_{F}$. The
intercepts of the straight lines with the energy axis give us the
threshold energies $E_{0,D}^{\textrm{semi-exp}}$ = 55.6 MeV and
$E_{0,F}^{\textrm{exp}}$ = 53.7 MeV.}
\end{figure}

In Fig.~\ref{s-factor}, we present the experimental $S_{F}(E)$ and
$S_{D}(E)$. For $S_{D}(E)$, use is made of
$\sigma_{D}^{\textrm{semi-exp}}$. From the zeros of $S_{i}(E)$, one
can deduce $E_{0,D}^{\textrm{semi-exp}}$=55.6~MeV and
$E_{0,F}^{\textrm{exp}}=$53.7~MeV. For both $i=F$ and $D$, the
observed $S_{i}$ are very well approximated by straight lines in the
subbarrier region and thus $E_{0,i}$ can be extracted without much
ambiguity. $E_{0,D}^{\textrm{semi-exp}}$ is found to be about 2 MeV
higher than $E_{0,F}^{\textrm{exp}}$, showing that the DR channels
open at higher energies than fusion channels, which is somewhat
unusual; normally the DR channels open at lower energies than fusion
channels. This unusual opening of the DR channels at higher energies
than fusion is related to the small DR cross sections at lower
energies as shown in Table I and Fig.~2. The $^{12}$C+$^{208}$Pb
system is a system in which DR takes place very weakly particularly
at lower energies.

$E_{0,i}$ may then be used as the energy where the imaginary
potential $W_{i}(E) (i=F,D)$ in Eqs.~(\ref{u_f}) and (\ref{u_d})
becomes zero, i.e., $W_{i}~(E_{0,i})=0$~\cite{kim1,kim2}. This
procedure will be used in the next subsection for obtaining a
mathematical expression for $W_{i}(E)$.

\subsection{$\chi^{2}$ Analyses}

All the $\chi^{2}$ analyses performed in the present work are
carried out by using the folding potential
described in IV.B
as the bare potential $V_{0}(r)$
and by using the fixed geometrical
parameters for the polarization potentials, $r_{F}$=1.40~fm,
$a_{F}$=0.35~fm, $r_{D}$=1.50~fm, and $a_{D}$=0.55~fm, which are
close to the values used in our previous study~\cite{kim1}. A slight
change of the values from those of Ref.~\cite{kim1} was made to
improve the $\chi^{2}$ fitting.

\begin{figure}
\begin{center}
\includegraphics[width=0.70\linewidth]{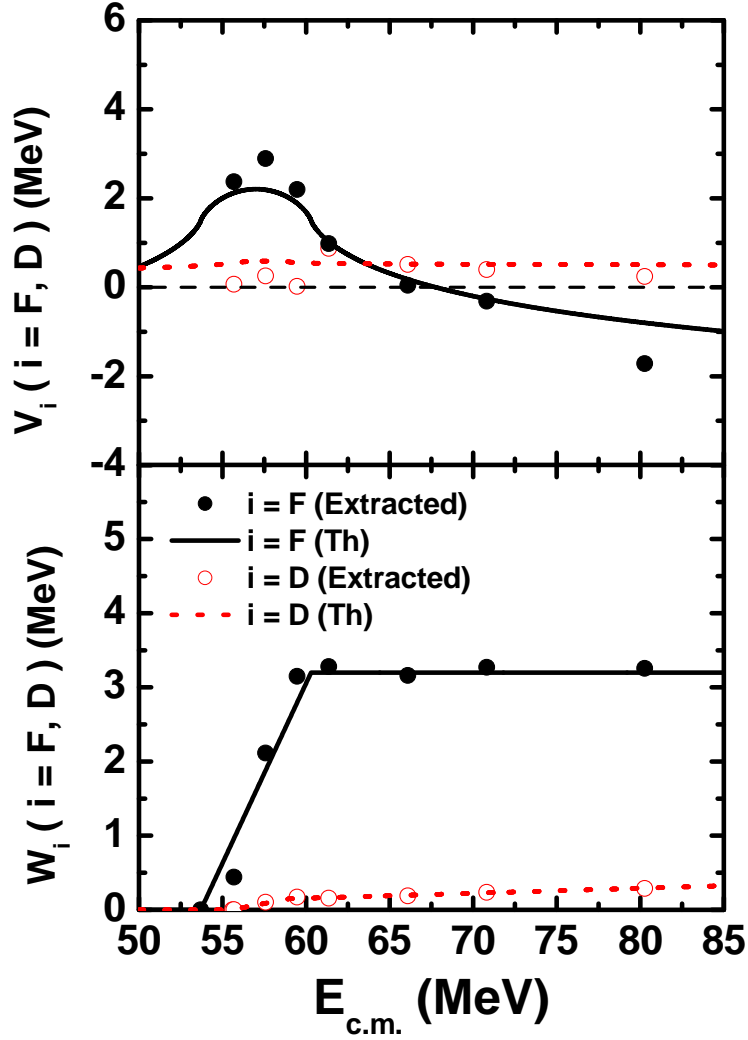}
\end{center}
\caption{\label{dispersion}~(Color online) The strength parameters
$V_{i}$ (upper panel) and $W_{i}$ (lower panel) for $i=D$ and $F$ as
functions of $E_{c.m.}$. The open and solid circles are the strength
parameters for $i=D$ and $F$, respectively. The dotted and solid
lines in the lower panel denote $W_{D}$ and $W_{F}$ from Eqs. (17)
and (18), respectively, while the dotted and solid curves in the
upper panel represent $V_{D}$ and $V_{F}$ calculated by using the
dispersion relation of Eq. (9) with $W_{i}$ given by Eqs. (17) and
(18). The values of $V_i (E_s )$ and the corresponding reference energies
$E_s$ used in Eq.~(9) are such that $V_F$ ($E_s$=60.3MeV) = $1.5$ MeV and
$V_D$ ($E_s$=59.0MeV) = $0.57$ MeV, respectively.}
\end{figure}

As in Ref.~\cite{kim1}, the $\chi^{2}$ analyses are done in two
steps; in the first step, all 4 strength parameters, $V_{F}(E)$,
$W_{F}(E)$, $V_{D}(E)$ and $W_{D}(E)$ are varied. In this step, we
can fix fairly well the strength parameters of the DR
potential, $V_{D}(E)$ and $W_{D}(E)$, in the sense that $V_{D}(E)$
and $W_{D}(E)$ are determined as a smooth function of $E$. The
values of $V_{D}(E)$ and $W_{D}(E)$ thus extracted are presented in
Fig.~\ref{dispersion} by open circles. The values of $W_{D}(E)$ thus
extracted can be well represented by the following function of $E$
(in units of MeV)
\begin{equation} \label{w_d}W_{D}(E) \; = \; \left \{
\begin{array}{lll}
0                  &\;\; \mbox{for $E\leq E_{0, D}^{\textrm{semi-exp}}=$55.6} \\
0.147(E-55.6)      &\;\; \mbox{for 55.6$<E\leq$59.0} \\
0.007(E-59.0)+0.15 &\;\; \mbox{for 59.0$<E\leq$111.0} \\
0.50               &\;\; \mbox{for 111.0$< E$} \\
\end{array}
\right. \vspace{2ex}
\end{equation}
Note that the threshold energy where $W_{D}(E)$ becomes zero is
set equal to $E_{0,D}^{\textrm{semi-exp}}$ as determined in the
previous subsection and are also indicated by the open circle at
$E=55.6$~MeV in Fig.~\ref{dispersion}. The dotted line in the lower
panel of Fig.~\ref{dispersion} represents Eq.~(\ref{w_d}). The dotted line in
the upper panel of Fig.~\ref{dispersion} denotes $V_{D}$ as
predicted by the dispersion relation of Eq.~(\ref{disper}), with
$W_{D}(E)$ given by Eq.~(\ref{w_d}). As seen, the dotted lines
reproduce the open circles fairly well, indicating that $V_{D}(E)$
and $W_{D}(E)$ extracted by the $\chi^{2}$ analyses satisfy the
dispersion relation.

In this first procedure of the $\chi^{2}$ fitting, however, the values of
$V_{F}(E)$ and $W_{F}(E)$ are not reliably fixed in the sense that
the extracted values fluctuate considerably as functions of $E$.
This is understandable from the expectation that the elastic
scattering can probe most accurately the optical potential in
the peripheral region, which is nothing but the region characterized
by the DR potential. The part of the nuclear potential responsible
for fusion is thus difficult to pin down in this first step.

To obtain more reliable information on $V_{F}$ and $W_{F}$,
we thus performed the second step of the $\chi^{2}$ analysis;
this time, instead of doing a 4-parameter search we fixed $V_{D}$
and $W_{D}$ as determined by the first $\chi^{2}$ fitting, i.e.,
$W_{D}(E)$ given by Eq.~(\ref{w_d}) and $V_{D}(E)$ predicted from
the dispersion relation. We then performed 2-parameter $\chi^{2}$
analyses, treating only $V_{F}(E)$ and $W_{F}(E)$ as adjustable
parameters. The parameter values thus determined are presented in
Fig.~\ref{dispersion} by solid circles.
The solid circles in the lower panel of Fig.~4 can be well represented by
\begin{equation} \label{w_f}
W_{F}(E) \; = \; \left \{ \begin{array}{lll}
0             &\;\; \mbox{for $E\leq E_{0, F}^{\textrm{exp}}=$53.7} \\
0.485(E-53.7) &\;\; \mbox{for 53.7$<E\leq$60.3} \\
3.20          &\;\; \mbox{for 60.3$< E$} \\
\end{array}
\right. \vspace{2ex}
\end{equation}
As is done for $W_{D}(E)$, the threshold energy where $W_{F}(E)$
becomes zero is set equal to $E_{0,F}^{\textrm{exp}}$ that is also
indicated by the solid circle in Fig.~\ref{dispersion}. As seen, the
$W_{F}(E)$ values determined by the second $\chi^{2}$ analyses can
fairly well be represented by the functions given by
Eq.~(\ref{w_f}).

Using $W_{F}(E)$ given by Eq.~(\ref{w_f}), one can generate
$V_{F}(E)$ from the dispersion relation. The resulting $V_F (r)$ is shown by
the solid curve in the upper panel of Fig.~\ref{dispersion}, which
again well reproduces the values extracted from the
$\chi^{2}$ fitting. This result shows that the fusion potential determined
from the present analysis also satisfies the dispersion relation.

Note that the energy variations in $W_{F}(E)$
and $V_{F}(E)$ are more rapid compared to those in $W_{D}(E)$
and $V_{D}(E)$, and are similar to those in tightly bound
projectiles~\cite{bae1,lil1,ful1}. It is thus seen that the
resultant $V_{F}(E)$ and $W_{F}(E)$ exhibit the threshold anomaly.

\subsection{Final Calculated Cross Sections in Comparison with
the Data}

\begin{figure}
\begin{center}
\includegraphics[width=0.80\linewidth]{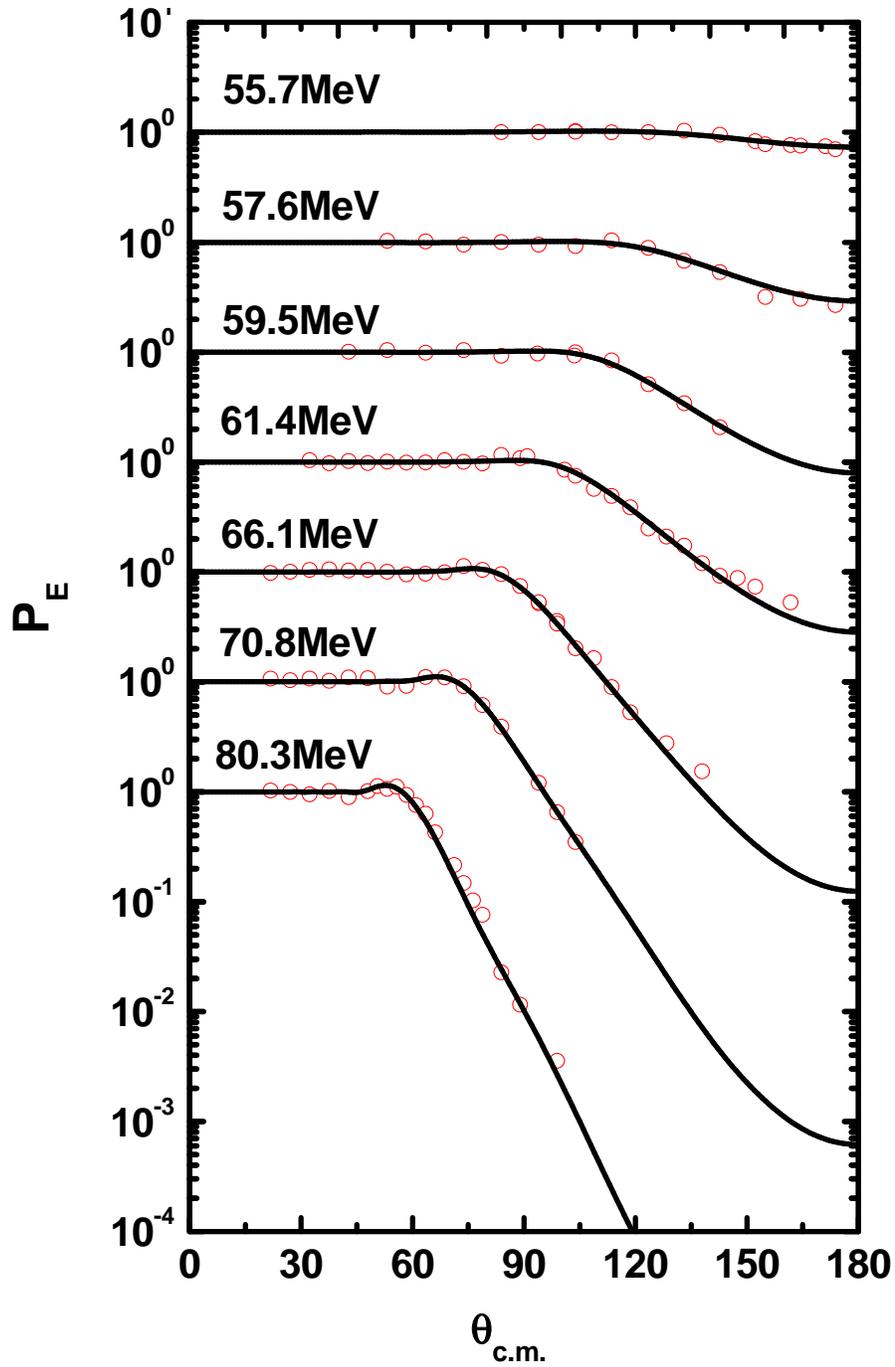}
\end{center}
\caption{\label{elastic}~(Color online) Ratios of the elastic
scattering cross sections to the Rutherford cross section calculated
with our final dispersive optical potentials are shown in comparison
with the experimental data. The data are taken from
Ref.~\cite{san1}.}
\end{figure}

\begin{figure}
\begin{center}
\vspace*{5.0cm}
\includegraphics[width=0.95\linewidth]{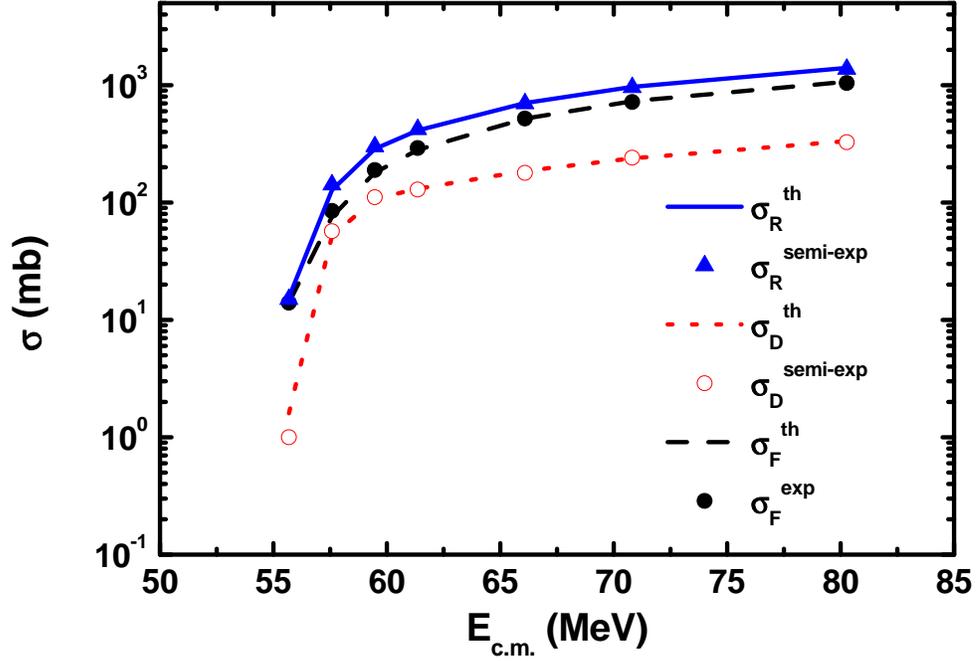}
\end{center}
\caption{\label{reaction}~(Color online) DR and fusion cross
sections calculated with our final dispersive optical potential are
shown in comparison with the experimental data.
$\sigma_{D}^{\textrm{semi-exp}}$ denoted by the open circles are
obtained as described in Sec.~III. The fusion data are from
Ref.~\cite{muk1}.}
\end{figure}

Using $W_{D}(E)$ given by Eq.~(\ref{w_d}) and $W_{F}(E)$ given by
Eq.~(\ref{w_f}) together with $V_{D}(E)$ and $V_{F}(E)$ generated
by the dispersion relation, we have performed the final
calculations of the elastic, DR and fusion cross sections. The
results are presented in Figs.~\ref{elastic} and~\ref{reaction} in
comparison with the experimental data. All the data are well
reproduced by the calculations.

It may be worth noting here that the theoretical fusion cross
section, $\sigma_{F}^{\textrm{th}}$, includes contributions from two
imaginary components $W_{I} (r)$ and $W_F (r) = W_{F}(E)f(X_{F})$ in
$U_{F}(r,E)$ of Eq.~(\ref{u_f}). In Table~\ref{fusion} the partial
contributions from $W_{I}(r)$ and $W_{F}(r)$, denoted by
$\sigma_{I}$ and $\sigma_{F}$, respectively, are presented
separately, together with the total theoretical fusion cross
section, $\sigma_{F}^{\textrm{th}}$. As seen, the contribution from
the inner part, $W_{I} (r)$, amounts to less than 10\% except at
highest energies $E_{c.m.}$ = 80.3~MeV, where the inner part
contributes by 14\%. This enhanced contribution from the inner part
at higher energies may be due to deeper penetration of the
projectile into the inner part at higher energies.

  It should be recalled at this stage that we assumed a constant
value of $W_{I}$=40 MeV. Such an assumption is apparently inconsistent with
a rapid energy variation expected to exist in the fusion potential
around the Coulomb barrier energy. Note, however, that elastic
scattering, fusion and total reaction cross sections are all rather
insensitive to the value of $W_{I}(r)$, in particular, at low
energies below the Coulomb barrier energy as discussed somewhat in details
in Sec IV.~B of Ref.~\cite{so2}.
Considering this and also the fact that $V_{I}(r)$,
the real potential associated with $W_{I}(r)$, would also be
insensitive to the observables, one could make the
inner part of the imaginary fusion potential $W_{I}(r)$ to be fully
dispersive and energy dependent.
We have not tried here to make such an
extension, since as emphasized earlier one cannot achieve it
without ambiguity due to the fact that
the observables cannot be reflective of the inner part of
the potential.

\begin{table}
\caption{Partial contributions $\sigma_{F}$ and $\sigma_{I}$ to the
fusion cross sections.} \vspace{2ex} \label{fusion}
\begin{ruledtabular}
\begin{tabular}{ccccc}
$E_{lab}$ (MeV) & $E_{c.m.}$ (MeV) & $\sigma_{I}$ (mb) & $\sigma_{F}$ (mb) &
$\sigma_{F}^{\textrm{th}}$ (mb) \\
\hline
58.9 & 55.7  &   1   &  13  &  14 \\
60.9 & 57.6  &   5   &  71  &  76 \\
62.9 & 59.5  &   9   & 171  & 180 \\
64.9 & 61.4  &  12   & 269  & 281 \\
69.9 & 66.1  &  37   & 482  & 520 \\
74.9 & 70.8  &  72   & 659  & 731 \\
84.9 & 80.3  & 149   & 925  &1073 \\
\end{tabular}
\end{ruledtabular}
\end{table}

\subsection{Discussions}

\begin{figure}
\begin{center}
\vspace{1.0cm}
\includegraphics[width=0.75\linewidth]{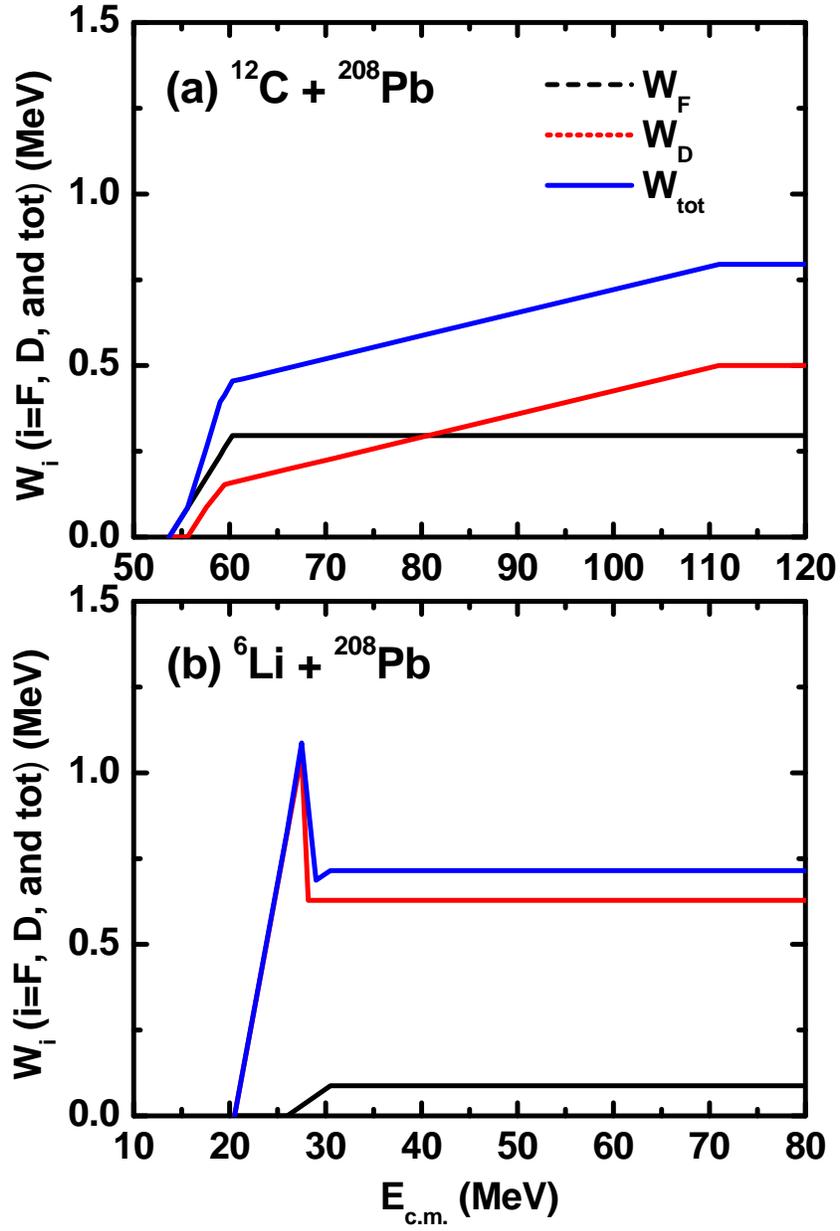}
\end{center}
\caption{\label{potential}~(Color online) The values of the fusion,
DR, and total imaginary potential, $W_{F}(r,E)$, $W_{D}(r,E)$ and
$W_{tot}(r,E)$, respectively, (a) at $r=R_{sa}=12.3$~fm for $^{12}$C
+ $^{208}$Pb and (b) at $r=R_{sa}=12.4$~fm for $^{6}$Li +
$^{208}$Pb.}
\end{figure}

 As already remarked in Sec. IV.D, the real and the imaginary parts of
both DR and fusion polarization potentials determined from the
present $\chi^{2}$ analyses satisfy the dispersion
relation~\cite{mah1,nag1} separately. We showed in Ref.~\cite{so1}
that for the case of the $^{6}$Li+$^{208}$Pb system the threshold
anomaly as was observed in heavy ion collisions involving
strongly bound projectiles~\cite{bae1,lil1,ful1} were distinctly seen only in
the fusion potential; the values of the DR potential changes with energy much
more slowly than those of the fusion potential. Now, for the present
case of $^{12}$C+$^{208}$Pb,
Fig.~4 shows that the values of $W_{D}(E)$
are smaller than those of $W_F (E)$ by about ten times.
However, a somewhat different
picture emerges if one plots the values of the imaginary parts of the DR
and fusion potentials at a strong absorption radius $r=R_{sa}$, i.e.,
$W_{D}(R_{sa},E)$ and $W_{F}(R_{sa},E)$, respectively.
In Fig.~\ref{potential} (a), plotted are the values of $W_{D}(R_{sa},E)$ and
$W_{F}(R_{sa},E)$, together with the sum, $W_{tot}(R_{sa},E)$, assuming
$R_{sa}=12.3$~fm.  In Fig.~\ref{potential} (b), we also
show for the sake of comparison the values of $W_D (R_{sa},E)$
and $W_F (R_{sa},E)$ at $r=R_{sa}=12.4$~fm for the $^{6}$Li+$^{208}$Pb system
obtained with the dispersive potentials of Ref.~\cite{so1}.

 It can be seen in Fig.~\ref{potential} (a) that the magnitudes of
$W_{D}(R_{sa},E)$ are now comparable with those of $W_{F}(R_{sa},E)$
and that $W_{D}(R_{sa},E)$ increases as much rapidly as
$W_{F}(R_{sa},E)$ does with almost the same threshold energies.
We can also see that
$W_{F}(R_{sa},E)$ is larger than $W_{D}(R_{sa},E)$ around
the Coulomb barrier energy for the $^{12}$C+$^{208}$Pb system,
which is consistent with the ratio $R_F$ shown in Fig.~2.
The sum $W_{tot}(R_{sa},E)$ of $W_{D}(R_{sa},E)$ and
$W_{F}(R_{sa},E)$ increases rapidly as typically observed for strongly
bound projectiles~\cite{bae1,lil1,ful1}.  In contrast to the above features
of $W_{tot}(R_{sa},E)$, $W_{D}(R_{sa},E)$, and $W_{F}(R_{sa},E)$ for
the $^{12}$C+$^{208}$Pb system, those of the $^{6}$Li+$^{208}$Pb
system are very different; first of all, as seen in
Fig.~\ref{potential} (b), $E_{0,D}^{\textrm{exp}} \ll
E_{0,F}^{\textrm{exp}}$ and the total $W_{tot}(R_{sa},E)$ is
dominated by $W_{D}(R_{sa},E)$,
which is again consistent with the ratio $R_F$ in Fig.~2,
and thus $W_{tot}(R_{sa},E) \simeq
W_{D}(R_{sa},E)$ over the whole energy range considered.

\section{Conclusions}

Simultaneous $\chi^{2}$ analyses are made for the elastic scattering
and fusion cross section data for the $^{12}$C+$^{208}$Pb system at
near-Coulomb-barrier energies based on the extended optical model
approach in which the polarization potential is decomposed into
direct reaction (DR) and fusion parts. Use is made of the double
folding potential as a bare potential. It is found that the
experimental elastic scattering and fusion data are well reproduced
without introducing any normalization factor for the double folding
potential and also that both DR and fusion parts of the polarization
potential determined from the $\chi^{2}$ analyses satisfy separately
the dispersion relation. Moreover, it is found that the imaginary
parts of both fusion and DR potentials at the strong absorption
radius show rapid energy variation around the Coulomb barrier energy
which is typical for tightly bound
projectiles~\cite{bae1,lil1,ful1}. The results are compared with
those for the $^{6}$Li+$^{208}$Pb system involving a loosely bound
projectile $^{6}$Li.

\acknowledgments The work was supported in part by the Korea Research
Foundation Grant funded by the Korean Government (MOEHRD)(KRF-
2006-214-C00014) and the Korea Science and Engineering Foundation
grant funded by the Korean Government (MOST)
(No. M20608520001-07B0852-00110).

\newpage

\newpage

\newpage

\newpage

\newpage

\newpage

\newpage

\end{document}